\documentclass[10pt,onecolumn]{IEEEtran}
\usepackage{graphicx}
\usepackage{psfrag}
\usepackage{subfigure}
\usepackage{overpic}
\usepackage{hyperref}
\usepackage{amssymb}
\usepackage{pgf}
\usepackage{pgffor}
\usepackage{tikz}
\usepackage{mathtools}
\usepackage{pst-sigsys}
\usepackage{stfloats}
\usepackage{cite}
\usetikzlibrary{patterns}
\usepackage{epsfig}
\usepackage{svg}
\usepackage{amsthm} 
\usepackage{amsfonts}
\usepackage{pgf}
\usepackage{pgffor}
\usepackage{tikz}
\usepackage{mathtools}
\usepackage{pst-sigsys}
\usepackage{stfloats}
\usepackage{graphicx}
\usepackage{cite}
\usetikzlibrary{patterns}
\usepackage{subfigure}
\usepackage{listings}
\usepackage{fp}
\usepackage{textcomp}
\usetikzlibrary{patterns,positioning,shapes,arrows,arrows.meta,graphs,svg.path,calc,fit}
\usepackage{pgfplots}
\usepackage{epsfig}
\usepackage{epstopdf}
\usepackage{booktabs}

\newcommand{\comment}[1]{}
\newcommand{\cC}{{\cal C}}

\newcommand{\beq}[1]{\begin{equation}\label{#1}}
\newcommand{\eeq}{\end{equation}}
\newcommand{\beqn}[1]{\begin{eqnarray}\label{#1}}
\newcommand{\eeqn}{\end{eqnarray}}

\newcommand{\Ddef}{\stackrel{\Delta}{=}}

\newtheorem{lemma}{Lemma}

\usepackage{filecontents}

\author{
    \IEEEauthorblockN{Marina Haikin\\}
    \IEEEauthorblockA  {
        Amazon, Tel Aviv, Israel \textsuperscript{\textsection}\\
        mkokotov@gmail.com\\
    }
    \and
   \IEEEauthorblockN{Ram Zamir\\}
    \IEEEauthorblockA  {
        Tel Aviv University, Israel\\
        zamir@eng.tau.ac.il
    }
}

\date{April 2021}

\title{Moments of Subsets of General Equiangular Tight Frames} 

\begin{document}

\maketitle
\begingroup\renewcommand\thefootnote{\textsection}
\footnotetext{Now at Amazon. Previously at Tel Aviv University where this work was performed.}
\endgroup

\begin{abstract}
    This note outlines the steps for proving that the moments of a randomly-selected subset of a general ETF (complex, with aspect ratio $0<\gamma<1$) converge to the corresponding MANOVA moments.
We bring here an extension for the proof of the 'Kesten-Mckay' moments (real ETF, $\gamma=1/2$) \cite{magsino2020kesten}. 
In particular, we establish a recursive computation of the $r$th moment,
for $r = 1,2,\ldots$,
and verify, using a symbolic program, that the recursion output coincides with the MANOVA moments.
\end{abstract}
\section{Recursive computation of moments}
Let $F$ be a unit-norm equiangular tight frame comprised of $n$ vectors in $\cC^m$, $\gamma=\frac{m}{n}<1$ is the aspect ratio of the frame. 
We define $s\Ddef (1-x)/2\sqrt{x}$, where $x = \gamma^{-1} - 1$
and denote a generalized conference matrix 
as 
\begin{equation}
\label{generalized_S}
 S =  \sqrt{\frac{n-1}{{x}}} \cdot \left( F' F - \frac{1}{2 \gamma} \cdot I_n \right)    
\end{equation}
which satisfies:
\begin{itemize}
    \item[(i)]
    $S_{i,i}$ = constant = $\sqrt{n-1} (1-x)/2\sqrt{x} \Ddef \sqrt{n-1}\cdot s$, 
    \item[(ii)]
    $S$ is conjugate-symmetric and $|S_{i,j}|=1$ for every $i,j\in[n]$ with $i\neq j$, and
    \item[(iii)]
    $S'S \propto I_n$, i.e., the off-diagonal elements of $S^2$ are zero, while the diagonal elements are 
equal to $(n-1) (x+1)^2/4x$.
\end{itemize}
Note that $x/(n-1)$ is the Welch bound, i.e., the squared absolute cross correlation between the frame vectors. For $\gamma=0.5$, 
we have
$x=1,s=0$,
so
$S =  \sqrt{n-1} \cdot \left( F' F - I_n \right)$, and $S'S=(n-1)I$, 
in agreement with the properties of the conference matrix in \cite{magsino2020kesten}.

We extend the proof in \cite{magsino2020kesten} to derive the expression for moments of the random submatrix $X_S = PSP$ defined as

\begin{equation}
\label{momentS}
    m^S_k=\frac{1}{n^{k/2+1}} \cdot 
E \left\{  {\rm trace} (X_S^k) \right\}=\sum_{t=1}^{k}\left( \sum_{\pi\in\prod (k,t)} V_n(\pi)\right)\cdot p^t
\end{equation}

where $P$ is a diagonal matrix such that the diagonal elements are independent random variables Bernoulli($p$).
For every partition $\pi$ of $[k]$ into $t$ blocks, $V_n(\pi)$ is sum of products of $k$ elements from the generalized conference matrix $S$, as a function of the partition. The sum is over all possible $t$ distinct values from $[n]$ according to the blocks of $\pi$ (value per block),
\begin{equation}
\label{V_n_pi}
    V_n(\pi)\Ddef \frac{1}{n^{k/2+1}}\sum_{a\in L_n(\pi)}S_{a(1),a(2)}S_{a(2),a(3)}\cdots S_{a(k),a(1)},
\end{equation}
\begin{equation}
\label{Ln}
    L_n(\pi)\Ddef\{a:[k]\rightarrow[n]:\{a^{-1}(a(i)):i\in[k]\}=\pi\} .
\end{equation}
The main difference from the KM case will be the computation of $V_n(\pi)$. Lemma 16 in \cite{magsino2020kesten} will be replaced by 
\begin{lemma}
\label{Lemma_V_n(pi)}
For every non-crossing partition $\pi \in \prod(k,t)$, the following holds: if the edges of $G_{\pi}$ partition into m simple cycles of sizes $l_1, \dots,l_m$, then as $n \rightarrow \infty$,
\begin{equation}
    V_n(\pi)\rightarrow A_{l_1}\cdots A_{l_m}
\end{equation}
where 
	\begin{equation}\label{A_recursion}
    A_{l+1} = -\sum_{i=1}^{l}A_i A_{l+1-i}, \,\,\, A_1 = s, A_2 = 1  .
\end{equation}
\end{lemma}
Note that for $\gamma=0.5$, $s=0$ and we get that for odd cycle, i.e., odd $l_i$, $A_{l_i}=0$, and for even $l_i$, $A_{l_i}=(-1)^{l_i/2-1} \cdot C_{l_i/2-1}$. Thus if $G_{\pi}$ contains any odd cycle, then $V_n(\pi)\rightarrow 0$ otherwise $V_n(\pi)$ converges to a product of Catalan numbers, in agreement with Lemma 16 in \cite{magsino2020kesten}.\\

For general $\gamma$ ($s\neq 0$), we don't have yet an a analogy to Lemma 8, but $ \sum_{\pi\in\prod (k,t)} V_n(\pi)$ can be computed by counting all $N(k,t)$ (Narayana number) non-crossing partitions of $[k]$ into $t$ blocks and for each induced cactus to compute $V_n(\pi)$ according to Lemma~\ref{Lemma_V_n(pi)}.

The desired moments of a (Bernoulli-$p$ selected) subset of ETF $F$ are then derived by using \eqref{generalized_S} and the binom formula:
\begin{align}
\label{binom}
m_k 
\nonumber
& \Ddef 
E \left\{ \frac{1}{n} \cdot {\rm trace} ((P F' F P)^k)  \right\}  
\\
\nonumber
&=
\sum_{j = 0}^k {k \choose j} \cdot  
x^{j/2}  \cdot \left( \frac{x+1}{2} \right)^{k-j}p^{j==0}  \cdot \frac{1}{n^{j/2+1}} \cdot 
E \left\{  {\rm trace} (X_S^j) \right\} 
\\
& =\left(\frac{x+1}{2} \right)^k\cdot p+
\sum_{j = 1}^k {k \choose j} \cdot  
x^{j/2}  \cdot \left( \frac{x+1}{2} \right)^{k-j}  \cdot m^S_j 
\end{align}
where $m^S_j$ is defined in \eqref{momentS}, and $j==0$ denotes $1$ if $j=0$ or $0$ otherwise.

Before we turn to the proof, we apply the recursion for several first moments:
\begin{alignat*}{5}
\nonumber
&  \,\,\,\,\,\, &&m^S_0 = 1 \,\,\,\,\,\, &&m_0 = 1\\
\nonumber
& A_1 =s \,\,\,\,\,\, &&m^S_1 = sp \,\,\,\,\,\, &&m_1 = p\\
\nonumber
& A_2 =1 \,\,\,\,\,\, &&m^S_2 = s^2p + p^2 \,\,\,\,\,\, &&m_2 = p+p^2x\\
\nonumber
& A_3 =-2s \,\,\,\,\,\, &&m^S_3 = s^3p + 3sp^2 + (-2s)p^3 \,\,\,\,\,\, &&\vdots\\
\nonumber
& A_4 =4s^2-1 \,\,\,\,\,\, &&m^S_4 = s^4p + 6s^2p^2 + (-8s^2+2)p^3 + (4s^2-1)p^4\\
\nonumber
& A_5 =-8s^3+6s \,\,\,\,\,\, &&\vdots\\
\nonumber
& A_6 =16s^4-24s^2+2
\end{alignat*}
%

%\rami{Excellent road map!!}

%%%%%%%%%%%%%%%%%%%%%%%%%%%%%%%%%%%%%%%%%%%%%%%%%%%%%%%%

\begin{proof}
We follow the notations in \cite{magsino2020kesten} $\Delta(a(1),a(2),\dots,a(k))=S_{a(1),a(2)}S_{a(2),a(3)}\cdots S_{a(k),a(1)}$.
If $a_j=a_{j+1}$ for any $j \in [k-1]$ or $a_k=a_1$, 

\begin{align}
    V_n(\pi)& =\frac{1}{n^{k/2+1}}\sum_{a\in L_n(\pi)}\Delta(a(1),\dots a(j-1),a(j+1),a(j+1),\dots,a(k))\label{v_pi_loop1}
    \\& =\frac{\sqrt{n-1}s}{n^{k/2+1}}\sum_{a\in L_n(\pi)}\Delta(a(1),\dots ,a(j-1),a(j+1),\dots,a(k))\label{v_pi_loop2}
    \\ & = \frac{s}{n^{(k-1)/2+1}}\sum_{a\in L_n(\pi\setminus \{j\})}\Delta(a(1),\dots ,a(j-1),a(j+1),\dots,a(k))+o(1)\label{v_pi_loop3}
\end{align}
were the first equality follows from the fact that the diagonal entries of $S$ are $s\sqrt{n-1}$, thus  $S_{a(j+1)a(j+1)}=s\sqrt{n-1}$. 
The restriction of $\pi\setminus\{j\}$ to $[k]\setminus{j}$ results in a partition $\pi'$ of $[k]\setminus{j}$ into same $t$ blocks besides removing $j$ from the block $\pi(j)$.
Thus the above expression of $V_n(\pi)$ implies 
\begin{equation}
    V_n(\pi)\rightarrow s\cdot V_n(\pi')  ,
\end{equation}
where throughout the proof, $\rightarrow$ means as $n \rightarrow \infty$.
This means that two consecutive indices which belong to same block and contribute a loop to the graph $G_{\pi}$, contribute a constant factor $s$ to $V_n(\pi)$.
This claim, as well as the derivation in \eqref{v_pi_loop1}-\eqref{v_pi_loop3}, extends for several loops. 
Given partition $\pi$ with $r$ loops in total (in one or more blocks), we can consider a \textit{squeezed} partition $\pi'$ with same number of blocks and $k'=k-r$. Each block $B'_i$ for $i=1,\dots,t$, is a squeezed block $B_i$, with only one representative index of sequence of several consecutive equal indices.
\begin{equation}
    V_n(\pi)\rightarrow s^r\cdot V_n(\pi')
\end{equation}
For example a partition 122234422333341 translated to 1234234 with a factor of $s^8$.

Now we focus on estimation of $V_n(\pi')$. $G_{\pi'}$ is a loop-free graph and thus Lemma 6 in \cite{magsino2020kesten} holds for the same reasons. Lemma 12(iv),12(v) also holds due to similar the properties of $S$, and thus Lemma 14 - which relies on all of the above - holds. 
This proves the claim of vanishing crossing partitions, 
i.e., let $\pi'\in \prod(k',t)$ be a crossing partition, 
then $V_n(\pi')\rightarrow 0$. 

We follow the proof by induction of Lemma 16 in \cite{magsino2020kesten}. Lemma 12(i) holds again by $S$ properties and thus for $t=2$, $V_n(\pi')=V_n(\{1\},\{2\})=1$.
As for $G_{\pi}$, with assumed $r$ cycles of sizes 1, proving the lemma for $V_n(\pi')$, will immediately imply the proof for $V_n(\pi)$, as $A_1^r = s^r$. 
Assuming that the (loops-removed) partition $\pi'$ is non-crossing, 
it follows that the graph
$G_{\pi'}$ is a cactus, 
and by Lemma 15 it is guaranteed to contain a singleton leaf (block in the partition). 
As we deal with $\pi'$, with $m'=m-r$ cycles, we aim to prove that 
\begin{equation}\label{Lemma16'}
    V_n(\pi')\rightarrow A_{l_1}\cdots\ A_{l_{m'}}.
\end{equation}

Case I of the proof refers to the case when the singleton block resides in a cycle of length 2, i.e., if the singleton is $\{k\}\in\pi$, $\pi(1)=\pi(k)$. A similar derivation as in \cite{magsino2020kesten} implies
\begin{equation}\label{case1}
    V_n(\pi')=V_n(\pi'')+o(1).
\end{equation}
Our induction hypothesis and \eqref{case1}, imply 
\begin{equation}
    V_n(\pi')\rightarrow A_{l_1}\cdots\ A_{l_{m'-1}}.
\end{equation}
Since $l_m=2$ and $A_2=1$, this establishes \eqref{Lemma16'}.

Case II of the proof refers to the case when $\{k\}\in\pi$ (the singleton block) resides in a cycle of length $l\ge3$. With similar analysis to \cite{magsino2020kesten} we get:
\begin{equation}\label{case2}
    V_n(\pi')=-\sum_{i=2}^{l}V_n({\pi'}^i)+o(1).
\end{equation}
Assuming $l_m=l$ and applying the induction hypothesis, we have
\begin{equation}
    V_n(\pi')\rightarrow-A_{l_1}\cdots\ A_{l_{m'-1}}\cdot  \sum_{i=2}^{l} A_{l-1}A_{l-i+1}=-A_{l_l}\cdots\ A_{l_{m'-1}}\cdot \sum_{i=1}^{l-1}A_{l}A_{l-i}.
\end{equation}
Applying the identity \eqref{A_recursion}, we get $V_n(\pi')\rightarrow A_{l_1}\cdots\ A_{l_{m'}}$, thereby establishing \eqref{Lemma16'}.
\end{proof}

%%%%%%%%%%%%%%%%%%%%%%%%%%

\section{IMPLEMENTATION OF THE DERIVED ALGORITHM}
\begin{lstlisting}[language=Python,basicstyle=\ttfamily\footnotesize]
import numpy as np
import sympy
from sympy.utilities.iterables import multiset_partitions
import scipy.special

# recursuve calculation of A(l) for l=1..n, arr_A[i]=A(i+1)
def calc_A(n,s):
    arr_A = np.zeros(n)
    arr_A[0] = s
    arr_A[1] = 1
    for i in np.arange(n-2)+2:
        arr_A[i] = -sum([arr_A[j]*arr_A[i-j-1] for j in np.arange(i)])
    return arr_A

# generation of all non crossing partitions of k to t blocks
def calc_partitions_k_t(k,t):
    partitions = list(multiset_partitions(np.arange(k),t))
    non_crossing = []
    non_crossing_len = []
    for part in partitions:
        cross = 0
        n = len(part)
        if n == 1:
            non_crossing.append(part)
            non_crossing_len.append([len(b) for b in part])
        else:
            for b1 in part:
                for b2 in part:
                    if (b2[0]>b1[0])&(len(np.array(np.where((b1>b2[0])&(b1<b2[-1])))[0])>0):
                        cross = 1
                        break
                if cross==1:
                    break
            if cross==0:
                non_crossing.append(part)
                non_crossing_len.append([len(b) for b in part])
    return non_crossing

'''
Find length of cactus leafs which correspond to a non-crossing partition
'''

# genrate lists off all intervals within blocks of partition
# list_pairs: a list of consecutive pairs in blocks [block[i],block[i+1]] for every block in the partition
# list_pairs_intervals: a list of intervals between consecutive pairs in blocks block[i+1]-block[i] for every
# block in the partition
def create_pairs_lists(part):
    list_pairs = []
    list_pairs_intervals = []
    for block in part:
        b_len = len(block)
        if b_len > 1:
            for i in range(b_len-1):
                list_pairs.append([block[i],block[i+1]])
                list_pairs_intervals.append(block[i+1]-block[i])
    return [list_pairs,list_pairs_intervals]

# sort the lists by the length of the intervals.
# start computing the the length of cycle which corresponds to an interval, from shorter to longer -
# for each interval substruct the cycles which correspond to smaller intervals contained in the interval
def update_lists(list_pairs,list_intervals):
    zipped_lists = zip(list_intervals, list_pairs)
    sorted_pairs = sorted(zipped_lists)
    tuples = zip(*sorted_pairs)
    list_intervals, list_pairs = [list(tuple) for tuple in tuples]
    for i in range(len(list_intervals)):
        intersections_length = 0
        for j in range(i):
            if (list_pairs[j][0] > list_pairs[i][0]) & (list_pairs[j][1] < list_pairs[i][1]):
                intersections_length = intersections_length + list_intervals[j]
        list_intervals[i] = list_intervals[i] - intersections_length
    return list_intervals

# generated the updated lists and add the cycle which correspond to an outer interval [k, 0],
# the length of which is k - sum of all cycles
def partition_to_cactus_sizes(part,k):
    [list_pairs,list_intervals] = create_pairs_lists(part)
    if len(list_pairs)>0:
        list_intervals = update_lists(list_pairs,list_intervals)
        cactus_sizes = list_intervals
        cactus_sizes.append(k - sum(cactus_sizes))
    else:
        cactus_sizes = [k]
    return cactus_sizes

'''
calculation of the 'centralized' moments
'''
# calculation of asymptotic V_n(pi) for a given partition by eq. 5 and sum over all possible (k,t) non-crossing
# partitions
def calc_V_pi_sum(k,t):
    V_pi_sum = 0
    cactuses = [partition_to_cactus_sizes(part,k) for part in calc_partitions_k_t(k,t)]
    for cactus in cactuses:
        V_pi_sum = V_pi_sum + np.prod(arr_A[np.array(cactus)-1])
    return V_pi_sum

# calculation of m_^s_k by eq. 2
def calc_m_s(k,p):
    m_s = 0
    for t in np.arange(k)+1:
        m_s = m_s + calc_V_pi_sum(k,t)*p**t
    return m_s


'''
binom formula for 'de-centralization' of the moments, eq.7
'''
def calc_m(k,x,p):
    m = ((x+1)/2)**k*p
    for j in np.arange(k)+1:
        m = m + scipy.special.binom(k,j)*x**(j/2)*((x+1)/2)**(k-j)*(m_s_vec[j-1])
    return m

# run calculation of 10 first moments
gamma = np.sqrt(0.5)
p = 0.6
x = 1/gamma - 1
s = (1-x)/(2*np.sqrt(x))
d = 10

arr_A = calc_A(d,s)
m_s_vec = [calc_m_s(k,p) for k in np.arange(d)+1]
m_vec = [calc_m(k,x,p) for k in np.arange(d)+1]
\end{lstlisting}
And the output is:
\begin{lstlisting}[basicstyle=\ttfamily\footnotesize]
------------------------------------
gamma = 0.7071067811865476
p = 0.6
------------------------------------
x = 0.4142135623730949
s = 0.45508986056222756
------------------------------------
calc ETF moments until d = 10
------------------------------------
A[l] for l = 1..10
[0.45508986056222756, 1.0, -0.9101797211244551, -0.1715728752538092, 1.9765215939999434, -2.2842712474619047,
 -2.1862701245144653, 10.126983722080917, -8.324794719173925, -21.347181155833745]
A[l] for l = 1..6 by manual application of eq.6
[0.45508986056222756, 1.0, -0.9101797211244551, -0.17157287525380915, 1.9765215939999432, -2.2842712474619047]
------------------------------------
example for all partitions (5,3) and the corresponding cactus lengths
[[0, 1, 2], [3], [4]]	[1, 1, 3]
[[0, 1, 3], [2], [4]]	[1, 2, 2]
[[0, 1], [2, 3], [4]]	[1, 1, 3]
[[0, 1, 4], [2], [3]]	[1, 3, 1]
[[0, 1], [2, 4], [3]]	[1, 2, 2]
[[0, 1], [2], [3, 4]]	[1, 1, 3]
[[0, 2, 3], [1], [4]]	[1, 2, 2]
[[0, 2, 4], [1], [3]]	[2, 2, 1]
[[0, 2], [1], [3, 4]]	[1, 2, 2]
[[0, 3], [1, 2], [4]]	[1, 2, 2]
[[0], [1, 2, 3], [4]]	[1, 1, 3]
[[0, 4], [1, 2], [3]]	[1, 3, 1]
[[0], [1, 2, 4], [3]]	[1, 2, 2]
[[0], [1, 2], [3, 4]]	[1, 1, 3]
[[0, 3, 4], [1], [2]]	[1, 3, 1]
[[0, 4], [1, 3], [2]]	[2, 2, 1]
[[0], [1, 3, 4], [2]]	[1, 2, 2]
[[0, 4], [1], [2, 3]]	[1, 3, 1]
[[0], [1, 4], [2, 3]]	[1, 2, 2]
[[0], [1], [2, 3, 4]]	[1, 1, 3]
example of cactus length for partition [[0,2],[1],[3],[4,5,7,10],[6],[8,9],[11]]
[1, 1, 2, 2, 2, 4]
------------------------------------
m_s[k] for k = 1..10
[0.2730539163373365, 0.4842640687119286, 0.35144954734733014, 0.5249702161277665, 0.44014600164558493,
 0.6031675926696468, 0.5443213610399722, 0.7080038062958746, 0.6683603744969959, 0.8395068505692707]
m_s[k] for l = 1..4 by manual application of eq.2
[0.2730539163373365, 0.4842640687119286, 0.35144954734733014, 0.5249702161277665]
------------------------------------
m[k] for k = 1..10
[0.6, 0.7491168824543142, 0.9949402589451768, 1.3553641553671059, 1.8704311672940084, 2.600699842184038,
 3.633212856926222, 5.091635541233709, 7.151079693455817, 10.05923853882072]
m[k] for l = 1..10 by Dubbs&Edelman (Manova moments triangles in Fig.1)
[0.6, 0.7491168824543142, 0.994940258945177, 1.355364155367106, 1.8704311672940082, 2.6006998421840386,
 3.633212856926222, 5.091635541233708, 7.1510796934558165, 10.059238538820706]
\end{lstlisting}
The output of a symbolic calculation is:
\begin{lstlisting}[basicstyle=\ttfamily\footnotesize]
------------------------------------
gamma = gamma
p = p
------------------------------------
x = x
s = s
------------------------------------
calc ETF moments until d = 10
------------------------------------
A[l] for l = 1..10
[s, 1, -2*s, 4*s**2 - 1, -8*s**3 + 6*s, 16*s**4 - 24*s**2 + 2,
 -32*s**5 + 80*s**3 - 20*s, 64*s**6 - 240*s**4 + 120*s**2 - 5, -128*s**7 + 672*s**5 - 560*s**3 + 70*s,
 256*s**8 - 1792*s**6 + 2240*s**4 - 560*s**2 + 14]
A[l] for l = 1..6 by manual application of eq.6
[s, 1, -2*s, 4*s**2 - 1, -8*s**3 + 6*s, 16*s**4 - 24*s**2 + 2]
------------------------------------
m_s[k] for k = 1..10 [to save space we show only first 6]
[p*s, p**2 + p*s**2, -2*p**3*s + 3*p**2*s + p*s**3,
 p**4*(4*s**2 - 1) + p**3*(-8*s**2 + 2) + 6*p**2*s**2 + p*s**4,
 p**5*(-8*s**3 + 6*s) + p**4*(5*s*(4*s**2 - 1) - 10*s) + p**3*(-20*s**3 + 10*s) + 10*p**2*s**3 + p*s**5,
 p**6*(16*s**4 - 24*s**2 + 2) + p**5*(36*s**2 + 6*s*(-8*s**3 + 6*s) - 6) +
 p**4*(15*s**2*(4*s**2 - 1) - 60*s**2 + 5) + p**3*(-40*s**4 + 30*s**2) + 15*p**2*s**4 + p*s**6, ...]
m_s[k] for l = 1..4 by manual application of eq.2
[p*s, p**2 + p*s**2, -2*p**3*s + 3*p**2*s + p*s**3,
 p**4*(4*s**2 - 1) + p**3*(-8*s**2 + 2) + 6*p**2*s**2 + p*s**4]


------------------------------------
gamma = gamma
p = p
------------------------------------
x = x
s = (-x + 1)/(2*sqrt(x))
------------------------------------
calc ETF moments until d = 10
------------------------------------
m[k] for k = 1..10
[0.5*p*(-x + 1) + p*(x/2 + 1/2),
 1.0*p*(-x + 1)*(x/2 + 1/2) + p*(x/2 + 1/2)**2 + 1.0*x**1.0*(p**2 + p*(-x + 1)**2/(4*x)),
 1.5*p*(-x + 1)*(x/2 + 1/2)**2 + p*(x/2 + 1/2)**3 + 3.0*x**1.0*(p**2 + p*(-x + 1)**2/(4*x))*(x/2 + 1/2) +
 1.0*x**1.5*(p**3*(x - 1)/sqrt(x) + 3*p**2*(-x + 1)/(2*sqrt(x)) + p*(-x + 1)**3/(8*x**(3/2))),
 2.0*p*(-x + 1)*(x/2 + 1/2)**3 + p*(x/2 + 1/2)**4 + 6.0*x**1.0*(p**2 + p*(-x + 1)**2/(4*x))*(x/2 + 1/2)**2 +
 4.0*x**1.5*(x/2 + 1/2)*(p**3*(x - 1)/sqrt(x) + 3*p**2*(-x + 1)/(2*sqrt(x)) + p*(-x + 1)**3/(8*x**(3/2))) +
 1.0*x**2.0*(p**4*(-x + (x - 1)**2)/x + p**3*(2 + 2*(-x + 1)*(x - 1)/x) + 3*p**2*(-x + 1)**2/(2*x) +
             p*(-x + 1)**4/(16*x**2)), ...]
m[k] for l = 1..10 by Dubbs&Edelman (Manova moments triangles in Fig.1)
[p,
 p**2*x + p,
 p**3*(x**2 - x) + 3*p**2*x + p,
 p**4*(x**3 - 3*x**2 + x) + p**3*(6*x**2 - 4*x) + 6*p**2*x + p, ...]
------------------------------------
difference between the simplified symbolic expressions of first 10 moments
[0, 0, 0, 0, 0, 0, 0, 0, 0, 0]
\end{lstlisting}

\bibliography{Moments_of_Subsets_of_General_Equiangular_Tight_Frames}
\bibliographystyle{IEEEtran}

\end{document}